\documentclass[twocolumn,draft,prd,nofootinbib]{revtex4}

\newif\ifpdf
\ifx\pdfoutput\undefined
\pdffalse 
\else
\pdfoutput=1 
\pdftrue
\fi

\ifpdf
\usepackage[pdftex]{graphicx}
\else
\usepackage{graphicx}
\fi

\usepackage{epsf}
\usepackage{dcolumn}

\newcommand{\be}{\begin{equation}}
\newcommand{\ee}{\end{equation}}
\newcommand{\Deln}{\ensuremath{\Delta N_\nu\;}}
\newcommand{\epm}{\ensuremath{e^{\pm}\;}}

\def\ie{{\it i.e.},~}
\def\eg{{\it e.g.},~}
\def\etal{{\it et al.}~}
\def\4he{$^4$He}
\def\3he{$^3$He}
\def\7li{$^7$Li}

\def\yd{$y_{\rm D}$~}

\def\Nnu{$N_{\nu}$~}

\newcommand\la{\lower0.6ex\vbox{\hbox{\ensuremath{\buildrel{\textstyle<}\over{\sim}\ }}}}
\newcommand\ga{\lower0.6ex\vbox{\hbox{\ensuremath{\buildrel{\textstyle>}\over{\sim}\ }}}}

\def\lsim{\mathrel{\raise.3ex\hbox{$<$\kern-.75em\lower1ex\hbox{$\sim$}}}}
\def\gsim{\mathrel{\raise.3ex\hbox{$>$\kern-.75em\lower1ex\hbox{$\sim$}}}}

\begin{document}
 \ifpdf
\DeclareGraphicsExtensions{.pdf,.jpg,.mps,.png}
 \else
\DeclareGraphicsExtensions{.eps,.ps}
 \fi


\title{Hiding relativistic degrees of freedom in the early universe}

\author{$^{1,7}$V. Barger, $^2$James P. Kneller, $^{3,7}$Paul Langacker,
$^{4,7}$Danny Marfatia and $^{5,6,7}$Gary Steigman}
\affiliation{$^1$Department of Physics, University of Wisconsin, Madison, WI 53706}
\affiliation{$^2$Department of Physics, North Carolina State University,
Raleigh, NC 27695}
\affiliation{$^3$Department of Physics and Astronomy, University of Pennsylvania,
Philadelphia, PA 19104}
\affiliation{$^4$Department of Physics, Boston University, Boston, MA 02215}
\affiliation{$^5$Department of Physics, The Ohio State University, Columbus,
             OH 43210}
\affiliation{$^6$Department of Astronomy, The Ohio State University, Columbus,
             OH 43210}
\affiliation{$^7$Kavli Institute for Theoretical Physics, University of
California, Santa Barbara, CA 93106}

\begin{abstract}

We quantify the extent to which extra relativistic energy density can be
concealed by a neutrino asymmetry without conflicting with the baryon
asymmetry measured by the Wilkinson Microwave Anisotropy Probe (WMAP).
In the presence of a large electron neutrino asymmetry, slightly more than
seven effective neutrinos are allowed by Big Bang Nucleosynthesis
(BBN) and WMAP at 2$\sigma$.
The same electron neutrino
degeneracy that reconciles the BBN prediction for the 
primordial helium abundance with the observationally
inferred value also reconciles the LSND neutrino
with BBN by suppressing its thermalization prior to BBN.

\end{abstract}

\maketitle

\section{Introduction}

If the three known neutrinos ($\nu_{e}$, $\nu_{\mu}$, $\nu_{\tau}$) mix
only with each other, all active neutrino degeneracies will equilibrate
to close to the electron neutrino degeneracy via neutrino oscillations
before BBN begins~\cite{equilibrate}.  There are no caveats to this
statement within the standard cosmological model for the oscillation
parameters relevant to atmospheric neutrino data and the Large Mixing
Angle (LMA) solution to the solar neutrino problem (which  has recently
been confirmed by the KamLAND experiment~\cite{kamland}). Thus, the
magnitude of the electron neutrino degeneracy allowed by BBN is of
special interest to any determination of, or constraints on, a lepton
asymmetry in the universe.

The authors of Ref.~\cite{equilibrate} emphasized that the degenerate
BBN scenario, in which a small $\nu_e$ asymmetry\footnote{For a neutrino
flavor $\alpha$, an asymmetry $L_\alpha$ between the numbers of
$\nu_\alpha$ and $\bar\nu_{\alpha}$ (``neutrino degeneracy'') can be
described by the neutrino chemical potential $\mu_\alpha$ or by the
dimensionless degeneracy parameter $\xi_\alpha \equiv \mu_\alpha/T$:
\be
L_\alpha\equiv {n_{\nu_{\alpha}}-n_{\bar\nu_{\alpha}} \over n_\gamma}=
{\pi^2 \over 12 \zeta(3)}\bigg(\xi_\alpha+{\xi_\alpha^3 \over \pi^2}\bigg)\,.
\ee
Note that $L_\alpha \approx 0.684 \xi_\alpha$ for $\xi_\alpha \ll
1$.}~$\xi_e \sim 0.2$ (which strongly affects the neutron to proton
ratio and, as a result, the primordial \4he abundance) is compensated
by a much larger $\nu_{\mu}$ or $\nu_{\tau}$ asymmetry $|\xi_{\mu,\tau}|
\sim 2-3$ (which affects the expansion rate) is excluded by this
equilibration, leading to a limit $|\xi_i| \lsim 0.1$ on all of the
asymmetries.

However, larger asymmetries may still be allowed if the effect of $\xi_e$
on BBN is compensated by a new source of energy density other than that
from the asymmetries themselves\footnote{The faster expansion rate due
to the new energy density could in principle also affect when or whether
the equilibration occurs. However, the considerations in this paper are
independent of equilibration.}.  In particular, if there are additional
degrees of freedom which do not mix with the three active neutrinos, but
which do contribute to the relativistic energy density at the BBN epoch,
their presence can be hidden by an electron neutrino degeneracy. In this
paper we quantify and explore the extent to which this statement is true.

As an example, light Dirac neutrinos involve three additional $SU(2)$-singlet
neutrinos $\nu_R$, the right-handed partners of the active states.  Within
the standard model (extended to include the right-handed neutrinos) these
states have no interactions other than those associated with neutrino mass,
\ie they can only be produced by mass effects of amplitude $m_\nu/E$ or the
associated tiny Yukawa couplings. These are too small to yield significant
number densities at the BBN epoch for masses in the sub-eV range.  However,
light Dirac neutrinos are especially motivated in models involving new gauge
interactions beyond the standard model (which may forbid the traditional
seesaw mechanism)~\cite{sbtalk}.  In that case they need not be sterile
with respect to the new interactions, and may be efficiently produced
prior to BBN~\cite{OSS}. This was recently studied in detail for a class
of models with $Z'$ couplings motivated by $E_6$~\cite{bll}. The production
and decoupling of the right-handed neutrinos, and the subsequent dilution
of their number density by annihilation of decoupled heavy particles and
color  confinement following the quark-hadron phase transition was calculated.
The results are dependent on the specific $Z'$ couplings and mass but,
typically, the energy density at BBN associated with the $\nu_R$ could
be equivalent to that carried by 3, 2, or 1 active neutrinos for a $Z'$
mass of 500 GeV, 1 TeV, or 2 TeV, respectively, with the constraints
becoming weak or disappearing for the specific couplings for which the
$Z'$ decouples from the $\nu_R$.

If sterile neutrinos that mix with the active ones exist, whether or
not flavor equilibration occurs is not known. We shall only consider
this case in the context of the LSND~\cite{lsnd} neutrino for which
a large electron neutrino degeneracy suppresses its mixing with the
active neutrinos~\cite{volkas}.  There has been considerable interest
in sterile neutrinos as a way to account for the antimuon neutrino
to antielectron neutrino oscillation signal observed in the LSND experiment
corresponding to a neutrino mass-squared difference scale of about
$0.1-1$ eV$^2$~\cite{lsndkarmen}.  With three neutrinos there are only
two independent mass-squared difference scales, that are fixed by the
observed atmospheric and solar oscillations to be about $2.5\times
10^{-3}$ eV$^2$~\cite{atm} and $7\times 10^{-5}$ eV$^2$~\cite{kam},
respectively.  Consequently, oscillations to a sterile neutrino have
been invoked to explain the higher mass-squared difference scale of
the LSND effect.  The MiniBooNE experiment~\cite{miniboone} now running
at Fermilab is designed to confirm or refute the LSND effect.  Analyses
with four neutrinos do not provide a good global fit to the data~\cite{valle},
but the possibility exists that the inclusion of certain small parameters,
hitherto neglected in the analysis, may improve the fit~\cite{weiler}.

We account for additional relativistic (at BBN) degrees of freedom
``$X$'' by normalizing their contribution to the energy density to
that of an equivalent, non-degenerate, neutrino flavor~\cite{ssg},
\be
\rho_{X} \equiv \Delta N_{\nu}\rho_{\nu} =
{7 \over 8}\Delta N_{\nu}\rho_{\gamma}.
\label{deln}
\ee
The effective number of neutrinos is $N_\nu \equiv 3 + \Delta N_{\nu}$.

Throughout this work the following primordial values are adopted for
the $^4$He mass fraction Y and for the deuterium abundance \yd $\equiv
10^{5}$(D/H):
\be
{\rm Y}= 0.238\pm 0.005\,, \ \ \ \  y_{\rm D}=2.6 \pm 0.4\,.
\ee
For a detailed discussion of these choices, see Ref.~\cite{pap}.

\section{The role of $\xi_e$ in BBN}

\begin{figure}[htbp]
\begin{center}
\epsfxsize=3.4in
\epsfbox{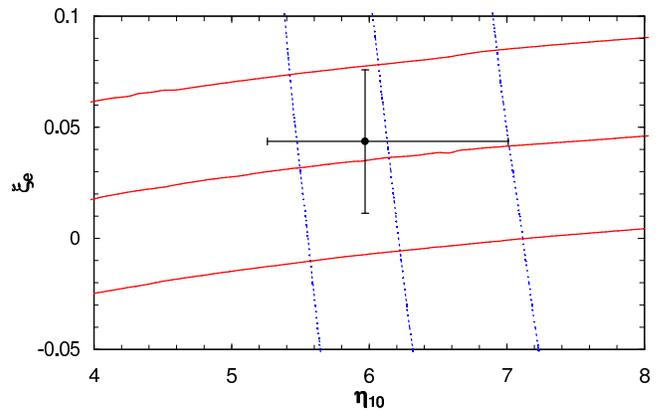}
      \caption{Isoabundance curves for D and \4he in the
      $\xi_{e}$ -- $\eta_{10}$ plane for \Nnu = 3.  The
      nearly horizontal curves are for  \4he (from top to
      bottom: Y = 0.23, 0.24, 0.25).  The nearly vertical
      curves are for D (from left to right: \yd$\equiv 10^{5}$(D/H)
      = 3.0, 2.5, 2.0).  The data point with error bars
      corresponds to \yd = $2.6\pm 0.4$ and Y = $0.238\pm
      0.005$.
      \label{f1}}

\end{center}
\end{figure}
The baryon-to-photon ratio $\eta$ provides a dimensionless
measure of the universal baryon asymmetry which is very small
($\eta\equiv n_{\rm B}/n_{\gamma}~\la 10^{-9}$)\footnote{We
define $\eta_{10} \equiv 10^{10}\eta$, to facilitate the use
of numbers of order unity.}.  By charge neutrality the asymmetry
in the charged leptons must also be of this order.  However, there
are no observational constraints, save those to be discussed here
(see \cite{ks,kssw} and further references therein), on the magnitude
of any asymmetry among the neutral leptons (neutrinos).  A relatively
small asymmetry between electron type neutrinos and antineutrinos
($|\xi_{e}| ~\ga 10^{-2}$), but large compared to the baryon asymmetry,
can have a significant impact on the early universe ratio of neutrons
to protons, thereby affecting the yields of the light nuclides formed
during BBN. The strongest effect is on the \4he abundance, which,
during BBN, is neutron limited.  For $\xi_{e} > 0$, there is an
excess of neutrinos ($\nu_{e}$) over antineutrinos ($\bar\nu_{e}$),
and the two body reactions regulating the neutron to proton ratio
($n + \nu_{e} \leftrightarrow p + e^{-}$, $p + \bar\nu_{e}
\leftrightarrow n + e^{+}$), drive down the neutron abundance;
vice-versa for $\xi_{e} < 0$.  The effect of a non-zero $\nu_{e}$ 
asymmetry on the relic abundances of the other light nuclides is 
much weaker.  This is illustrated in Fig.~\ref{f1}, which shows 
the D and \4he isoabundance curves in the $\xi_{e} - \eta_{10}$
plane for \Nnu = 3.  The nearly horizontal curves reflect the weak
dependence of Y on the baryon density, along with its significant
dependence on the neutrino asymmetry.  In contrast, the nearly vertical
curves reveal the strong dependence of \yd on the baryon density and
its weak dependence on any neutrino asymmetry.  This complementarity
between \yd and Y permits the pair \{$\eta,\Delta N_{\nu}$\} or the
pair \{$\eta,\xi_{e}$\} to be determined once the primordial abundances
of D and \4he are inferred from the appropriate observational data.

\section{Analysis}

\begin{figure}[htbp]
\begin{center}
\epsfxsize=3.4in
\epsfbox{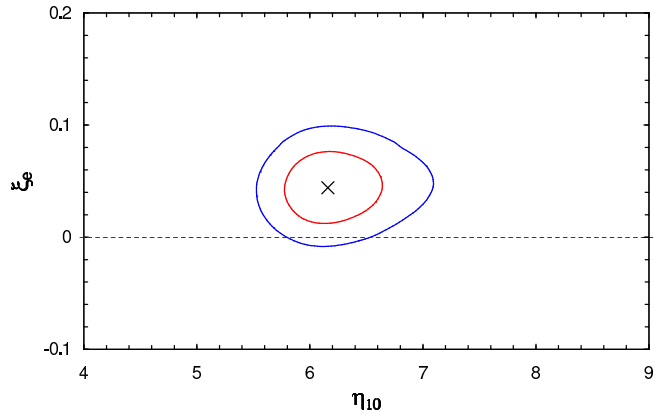}
      \caption{The $1\sigma$ and $2\sigma$
               contours in the $\eta_{10} - \xi_e$ plane
               for \Nnu = 3 from a joint CBR -- BBN fit
               using WMAP data and the adopted D and \4he
               abundances. The cross marks the  best-fit
               point $(\eta_{10},\xi_e)=(6.16,0.044)$.
      \label{f2}}
\end{center}
\end{figure}

In the ``standard'' case of \Nnu = 3, the observed abundances of D
and \4he may be used to identify the allowed region in the $\eta - 
\xi_{e}$ plane.  In addition, the cosmic background radiation (CBR) 
data from WMAP~\cite{map}, 
provide complimentary information for the $\eta$ distribution 
(for different \Nnu\cite{pap}),
which will further restrict the allowed $\eta$ range.  In Fig.~\ref{f2} 
are shown the $1\sigma$ and $2\sigma$ contours in the $\eta_{10} - 
\xi_e$ plane for \Nnu = 3 from a joint CBR (WMAP) -- BBN fit using the
adopted D and \4he abundances.  Although very large when compared to
the baryon asymmetry, the allowed values of $\xi_{e}$ are sufficiently
small ($|\xi_{e}| \lsim 0.1$) that the ``extra'' energy density contained
in such neutrinos is negligible~\footnote{$\Delta N_{\nu}(\xi) = \sum_i \big[{30
\over 7}({\xi_i \over \pi})^{2} + {15 \over 7}({\xi_i \over \pi})^{4}\big] ~\la
0.013$; see, \eg \cite{ks}.}, justifying the claim that these results
do, in fact, represent \Nnu = 3.

\vskip 0.1in
\vskip 0.1in

\subsection{The LSND neutrino}

If the LSND results~\cite{lsnd} are to be accounted for by the mixing of a
``sterile" neutrino with one of the active neutrinos, it is of importance
to consider whether the LSND neutrino will be populated, through mixing
with the active neutrinos, sufficiently early in the evolution of the
universe that it contributes \Deln = 1 to the relativistic energy density
prior to BBN.  Given the parameters required by LSND, in the absence of
any significant asymmetry among the active neutrinos the LSND neutrino would,
indeed, have come into thermal equilibrium sufficiently early~\cite{volkas}.
In this case the new ``standard" model would have \Nnu = 4.  However, in
the absence of a significant neutrino degeneracy, \Nnu = 4 is strongly
excluded~\cite{pap}.

In contrast, as the asymmetry in the active neutrinos increases, the
sterile-active mixing is {\it delayed}~\cite{volkas}.  For $\xi_{e}
~\ga 0.01 - 0.1$, mixing occurs {\it after} the active neutrinos have
decoupled from the \epm -- $\gamma$ plasma.  In this case, although
the sterile neutrino does still mix with the active ones, the energy
is shared among all the neutrinos and \Nnu remains equal to 3.  It is
interesting, even amusing, that the range of $\xi_{e}$ required by BBN
for \Nnu = 3 is, in fact, in excellent agreement with that required to
{\it delay} the thermalization of the LSND neutrino ($\xi_e \gsim 0.01$
-- 0.1), thereby keeping \Nnu $\sim 3$. Thus, the same values of $\xi_{e}$
that reconcile the low value of Y with the BBN prediction (for \Nnu = 3)
also reconcile the LSND sterile neutrino with BBN by delaying its mixing
with the active neutrinos.

\subsection{Arbitrary \Nnu}

\begin{figure}[htbp]
\begin{center}
\epsfxsize=3.4in
\epsfbox{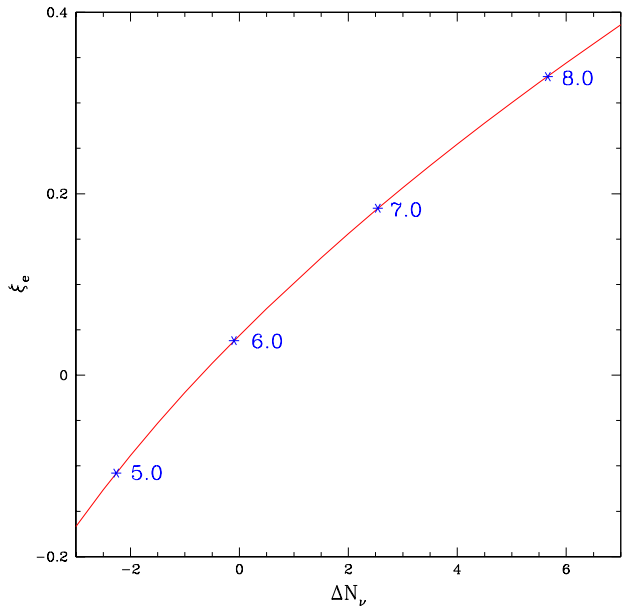}
      \caption{The approximate $\xi_{e}$ -- \Deln relation
               corresponding to \yd = 2.6 and Y = 0.238.
               The numbers shown are the corresponding values
               of $\eta_{10}$ at those points.
      \label{f4}}
\end{center}
\end{figure}

Let us set aside the LSND neutrino, assume that \Nnu = 4, and ask if BBN 
and the adopted primordial abundances can be reconciled (along with the 
CBR) through the appropriate choice of a nonzero asymmetry in the active
neutrinos.  

\begin{figure}[htbp]
\begin{center}
\epsfxsize=3.4in
\epsfbox{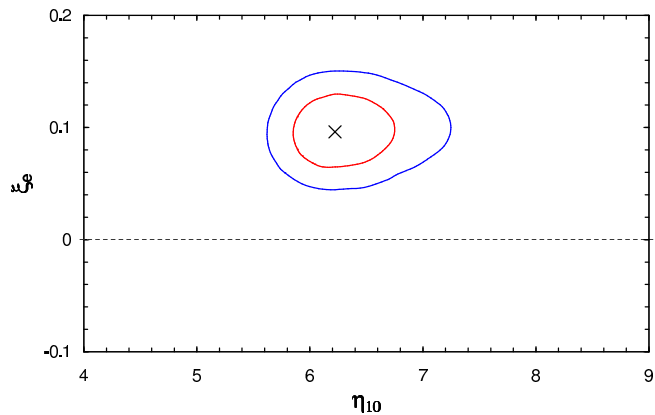}
      \caption{Same as Fig.~\ref{f2}, except \Nnu = 4.
               The cross marks the best-fit point 
               $(\eta_{10}, \xi_e)=(6.22, 0.096)$.
      \label{f3}}
\end{center}
\end{figure}

From  BBN with fixed abundances of D and \4he, as \Nnu increases, so too 
will the best fit values of $\eta$ and $\xi_{e}$.  It is straightforward 
to understand this effect.  As \Nnu increases, the early universe expands 
more rapidly, leaving {\it less} time to burn D.  As a result, for a 
{\it fixed} baryon to photon ratio, \yd {\it increases} with increasing 
\Nnu~\footnote{This effect on the primordial deuterium abundance is often, 
but not always~\cite{cfo,pap}, ignored.  However, for $|\Delta N_{\nu}| 
~\ga 1$, it is not negligible.}.  To reduce \yd back towards its observed 
value requires increasing the baryon density.  The combination of an 
increased baryon density and \Nnu = 4, raises the predicted primordial 
abundance of \4he, requiring a larger $\xi_{e}$ to reconcile the BBN 
predictions with the data.  Indeed, it is clear from Fig.~1 of Kneller 
\etal \cite{kssw} that for {\it any} (reasonable) choice of \Nnu there 
is a pair of $\{\eta_{10}^{*},\xi_{e}^{*}\}$ values for which {\it perfect}
agreement with the observed D and \4he abundances can be obtained
($\chi^{2}_{\rm BBN} = 0$). An approximation to this $\xi_{e}$
-- \Deln relation~\cite{bbnped} is shown in Fig.~\ref{f4} for
\yd = 2.6 and Y = 0.238; the numerical values along the curve are the
corresponding values of $\eta_{10}$. However, this correlation of increasing
$\eta$ with increasing \Nnu is broken by the inclusion of WMAP data
through the CBR-imposed constraint on the $\eta$ -- \Nnu relation~\cite{pap}.
As a result, while for \Nnu = 4 the allowed values of $\xi_{e}$ increase
from that for \Nnu = 3, the shift in $\eta$ is much more limited.  In 
Fig.~\ref{f3} is shown the analogue of Fig.~\ref{f2} (the confidence 
contours in the $\eta_{10} - \xi_e$ plane) for \Nnu = 4.

\begin{figure}[htbp]
\begin{center}
\epsfxsize=3.4in
\epsfbox{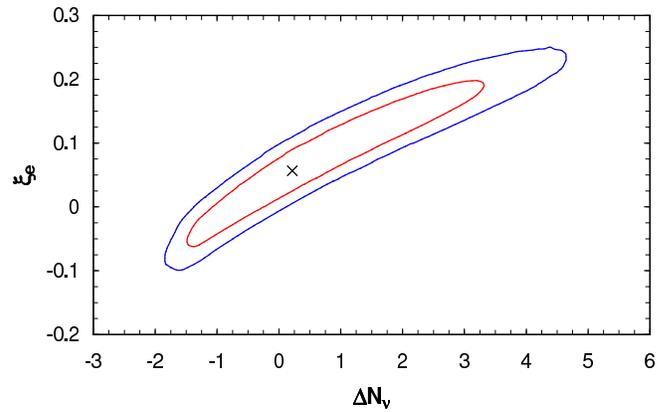}
      \caption{The $1\sigma$ and $2\sigma$ contours in the
               \Deln -- $\xi_{e}$ plane using WMAP data and
               BBN with the adopted D and \4he abundances.
               The cross marks the best-fit point $(\Delta N_\nu, \xi_e) =
               (0.22, 0.057)$.
                    \label{f5}}

\end{center}
\end{figure}

It is clear that from BBN alone, reconciling the primordial D and \4he 
abundances with the BBN predictions for large \Nnu would require too 
large a baryon to photon ratio to be consistent with the $\eta$ -- \Nnu 
constraints from WMAP.  The parameter pair \{$\Delta N_{\nu},\xi_{e}$\} 
can therefore be constrained using the \{$\Delta N_{\nu},\eta$\} likelihood 
distribution obtained~\cite{pap} from the WMAP data~\cite{map} along with 
the \{$\xi_{e},\Delta N_{\nu},\eta$\} likelihood distribution from our 
BBN analysis, by forming the joint likelihood and marginalizing over 
$\eta$ with the prior $\pi(\eta)$.  Adopting a flat p.d.f. for $\pi(\eta)$ 
the \{$\Delta N_{\nu},\xi_{e}$\} constraints are derived after calculating 
(up to a multiplicative constant),
\be
L(\xi_{e},\Delta N_{\nu}) = \int L_{\rm BBN}(\xi_{e},\Delta N_{\nu}, \eta)
\times L_{\rm WMAP}(\Delta N_{\nu},\eta) d\eta.
\ee
This analysis results in the allowed regions in the \Deln -- $\xi_{e}$ 
plane shown in Fig.~\ref{f5}, providing limits on the values of \Deln 
that can be accommodated by non-zero $\xi_e$ without conflicting with 
the $\eta$ and \Deln values allowed by WMAP.

\section{Conclusions}

Big Bang Nucleosynthesis places strong constraints on any new source of
relativistic energy density present during the BBN epoch.  Examples include
the energy density associated with large asymmetries $|\xi_{\mu,\tau}|$
between $\mu$ or $\tau$ neutrinos and their antineutrinos, with the
right-handed components $\nu_R$ of light Dirac neutrinos if they can be
efficiently produced by new interactions, or with sterile neutrinos
which mix with active neutrinos, as suggested by the LSND experiment.
In the absence of a significant neutrino asymmetry the data constrain
the equivalent number of new neutrino flavors \Deln to $\Delta N_{\nu}
< 0.1-0.3$ (assuming that $\Delta N_{\nu} \ge 0$)~\cite{pap}.

BBN also constrains the electron neutrino asymmetry, $\xi_e$, which
directly affects the neutron to proton ratio and therefore the \4he
abundance.  For $\Delta N_{\nu}=0$ we find $\xi_e \lsim 0.1$, as
shown in Fig.~\ref{f2}. However, there remains the possibility
of compensation between the effects of $\Delta N_{\nu}$ and $\xi_e$,
which can relax the constraints on both.  The expanded ranges,
consistent with WMAP and BBN, are shown in Fig.~\ref{f5}.  For $\xi_{e}$
in the range $-0.1 ~\la \xi_{e} ~\la 0.3$, values of \Deln in the
range $-2 ~\la \Delta N_{\nu} ~\la 5$ are permitted.  After marginalizing 
over $\xi_{e}$, the $2\sigma$ range  
for \Deln (obtained by identifying those values of \Deln for which the 
likelihood
is above $L_{max}/e^2$) is $-1.7 ~\la \Delta N_{\nu} ~\la 4.1$.

The possibility of compensation between a small $\xi_e$ and a large 
$|\xi_{\mu,\tau}|$ is now excluded given the oscillation parameters 
inferred for the atmospheric neutrinos and for the LMA solution for 
the solar neutrinos because of the equilibration of the three 
$\xi_i$~\cite{equilibrate}.  However, compensation between $\xi_e$ 
and other types of relativistic energy density, such as neutrinos 
which do not mix with active ones, remains a possibility.  We have 
explored this possibility in detail in this paper finding, for example, 
that $\Delta N_{\nu}=1$ is consistent with the BBN and the adopted 
primordial abundances provided that $\xi_e \sim 0.1$, as shown in 
Fig.~\ref{f3}.  Another consequence of our analysis is that even 
three, light, Dirac neutrinos ($\Delta N_{\nu}=3$) can be accomodated 
provided that $\xi_e \sim 0.2$; see Fig.~\ref{f5}.

Mixing between sterile and active neutrinos must be considered
separately, because $\xi_e$ can shift the time/temperature at
which the sterile neutrinos are produced~\cite{volkas}.  Values
of $\xi_e$ in the range $0.01-0.1$ (which are allowed for $\Delta
N_{\nu} \sim 0$; see Fig.~\ref{f2}), are sufficient to keep
\Nnu = 3 and thus reconcile the LSND neutrino with BBN.

The compensation between \Deln and $\xi_e$ explored here is, perhaps,
fine-tuned.  It should not be dismissed, however, because of the 
significance of these constraints for particle physics (in particular, 
for models of neutrino mass and mixing), and also because of the 
importance of $\Delta N_{\nu}$ for the subsequent evolution of 
structure in the universe, as probed by the CMB and large scale
structure surveys.  Also, a large $\xi_e$ of the magnitude relevant
for BBN (\eg $\xi_e \sim {\rm few} \ \times 10^{-1}$) is huge
compared to the baryon asymmetry, so even if it is unlikely, a
value in this range would be of profound significance for particle
physics and cosmology.  While in many models of leptogenesis the
asymmetry among the neutral leptons is tied to that in the charged
leptons and, hence, is very small ($\xi_{e} \approx \eta \sim 10^{-9}$),
there are mechanisms for generating the large asymmetry required
here~\cite{largeasym}.

\acknowledgments
We thank K.~Abazajian, G.~Fuller and Y.~Wong for enlightening
discussions.  This research was supported by the U.S.~DOE under
Grants No.~DE-FG02-95ER40896, No.~DE-FG02-91ER40676, No.
DE-FG02-91ER40690, DE-FG02-02ER41216, and No. DOE-EY-76-02-3071,
by the NSF under Grant No.~PHY99-07949, and by the Wisconsin
Alumni Research Foundation. VB, PL, DM and GS thank the Kavli
Institute for Theoretical Physics at the University of California,
Santa Barbara for its support and hospitality.

\end{document}